# PROSPECTS OF USING ANTAGONIST HISTAMINE H2-RECEPTOR (CIMETIDINUM) AS ADJUVANT FOR MELANOMA BIOTHERAPY TREATMENT.

**Authors:** *I.V. Manina, N.M. Peretolchina, N.S. Saprikina, A.M. Kozlov, I.N. Mikhaylova, K.I. Jordanya A.Y. Barishnikov.*

N.N. Blokhin Russian Cancer Research Center of RAMS, Moscow, Russian Federation.

\* **Correspondence should be addressed to:**

Irina Manina,
N.N. Blokhin Russian Cancer Research Center, Kashirskoye shosse, 24, Moscow, Russia, 115478

**e-mail:** ira-bio@yandex.ru, irina.v.manina@gmail.com

## Summary

Improvement of anti-tumor biotherapy effectiveness by modification of immune response with histamine H2 receptor Cimetidinum (CM) was studied using the experimental murine model of B16 F10 melanoma *in vivo*. It is shown that skin melanoma biotherapy by antitumor whole-cell GM-CSF-producing vaccine with the addition of CM (in dose of 25 mg/kg, daily for 5 days) increases preventive effects of vaccination. 33 % of mice did not have tumor growth within 60 days of observation. Average life span of animals exceeded those of the control group up to 68 %. Using CM-combined bio-chemotherapy doesn't improve therapeutic effect, however in the case of a monotherapeutic approach tendency for increased average life-time and decreased metastatic processes in mice with the developed tumors was noticed. Aquired data provides expediency to study the further application of CM as adjuvant for skin melanoma vaccinotherapy, and also the necessity to verify the immune status of an organism against the complex bio-chemotherapy.

INTRODUCTION

Changing of Th1/Th2 equilibrium leads to significant changes of immune reactions in patients with skin melanoma [1,2]. The patients with metastatic melanoma of 3-4 stages have the Th2 immune chronic inflammation. The increase of IL 4,5,6,9,10 cytokine and histamine concentrations result in disease progression and to an increase in metastatic process [5,6,7,8,9,10,11,12]; while Th1 immune profile correlates with propitious forecast of the disease [5,6,7]. It gives us an opportunity to model various immune reactions for increasing the efficacy of skin melanoma therapy utilizing concept approaches of allergooncology [13].

While describing histamine importance in the immune reactions it is necessary to focus our attention on following [14,15]: 1) lymphocytes express 2 types of histamine receptors: H1-R, H2-R; 2) histamine interaction with G-protein coupled H1-R leads to activation of immune reaction by means of T-cell or B-cell receptors; 3) histamine action through receptor H2-R is realized by adenylatecyclase, and leads to subsequent elevation of intercellular cAMP levels and to reversing immunosuppressive effect; 4) H1-R is expressed exclusively by Th1 cells; Th2 cells have both H1-R and H2-R on their surface. Besides histamine is characterized by the stimulatory effect on cytotoxic CD8+ T lymphocytes and the suppressive H2-R-mediated influence on the NK; 5) histamine effect on different population of lymphocytes can be realized by monocytes which react to it by increased production of immunoregulatory IL18 and IFN alongside with a decrease in producing of active forms of oxygen and phagocytic activity; 6) histamine effect on macrophages cancels of the inhibitory macrophage effects through active forms of oxygen on T lymphocytes and NK cells; 7) endogenous histamine influences the expression of surface receptors and adhesion molecules in immune cells [16,17]. Histamine takes part in inflammation and reparation processes by increasing vascular penetration; activates cytokine cascade and stimulates immune cells function, stroma cells function and stimulates angiogenesis. As for oncogenesis, histamine can stimulate proliferation and tumor angiogenesis process increasing tumor growth [18].

There are histamine receptors on the surface of both normal cells and tumor cells where their biological functions are not clear so far. In same investigations it was shown that histamine does not only stimulate tumor cell proliferation, but is also chemotoxic to both human cancer (Hela and A431) and melanoma (A875) containing H1-R [19].

Considering that the effects caused by histamine can be blocked by H-R antagonists, in 1970-1980 the influence of H-R antagonists as pharmacological instruments of influence on the types and kinetics of immune reactions were studied. Cimetidinum (CM) was used as a standard drug. Immune-regulating potencies of H2-R antagonists have the following order: Cimetidinum> Ranitidinum>Famotidinum. CM is able to directly block H2-R and indirectly block H1-R [20,21].

Applying CM for 7 days *per os* to healthy donors resulted in leukocytosis, neutrophilia, increase in CD3+ Tl population CD4+ cells. The number of NK cells decreased while the number of NKT cells did not change [22]. Small dozes of histamine H2-inhibitors stimulate sIgA synthesis and thus activating "macrophages-fibroblasts" system [23,24].

CM realizes its activity by blocking tumor angiogenesis and retarding cancer cell proliferation. CM also realizes its activity by increasing lymphocytes activity and in such a way it decreases the function of Tl suppressors. It increases antitumor immune reaction of the whole organism by retarding histamine activity as of a tumor growth factor, by increasing IL2 level, and Ig production [25,26]. CM lowers skin EGF (epidermis growth factor) by decreasing intracellular cATP level [27].

CM counteracts IL-1 inducted activation of adhesion molecules in endothelial vascular cells as well as in liver sinusoids. CM significantly decreases A-selectin protein expression without altering its at mRNA expression levels in cells, which can prevent hematogenic metastasis [28]. However, it is shown that after CM cessation the E-selectin level increases again [29].

Antitumor and immune regulating CM activity was first studied in patients with GI cancers. Taking into consideration good results from these studies, the similar investigations of CM functions were undertaken on tumors of other tissues, which contained a big number of histamine producing cells.

CM by means of activating immune system cells and increasing antitumor T-cells response improved the prognosis and increased GI cancer patients survivability after surgical treatment as well as in combination with chemical therapy of non-operable colonic and rectum 4 stage cancers [17,30].

CM is able to affect on caspase cascade activation in vitro, including 8, 9, 3 caspases. It is shown that, in this case, expression of Bcl protein is decreased while Bac protein expression is increased. Thus CM is able to induce apoptosis in tumor cells [31].

Experiments on SCID mice treated with CM it have shown growth inhibition of pancreas tumor cells xenotransplantants [32,33]. CM significantly reduced the development of metastases and increased life span of mice with neoblastoma and glioma during the experiment [34,35,36].

It was also shown that CM is able to slightly increase TCGF production after PHA-P activation. More of all, CM significantly increased TCGF-activated PBL protein proliferation. This effect was demonstrated in patients with skin melanoma who had initially low Tl activity.

Skin and melanoma cells have significant levels of histamine and express histamine receptors, indicating on the existence of autocrine and paracrine histamine regulation. In melanoma cells histamine increases intracellular cATP concentration and stimulates melanoma cell growth and metastasis progress [37,38]. It is demonstrated that histamine increases melanoma cell growth mainly through H2-R. Tumor progression decreases in SCID mice with HT168 human melanoma xanotransplantant melanoma cell line in vivo with combined CM and tamoxifen derivative therapy regulating P453 cytochrome (DPPE). Such regulations are associated with an increase in gIFN levels and an increase of infiltration of tumor tissue by macrophages. Thus, there are 2 active mechanisms: direct inhibition of tumor cell proliferation by H2-R antagonists and immune reaction activation (Th1 type), characterized by IFN-production [39,40].

In melanoma cells histamine stimulates human Ets-2 proto-oncogene synthesis via H2-R [41]. It is shown that H2-R histamine antagonists differing from H1-R histamine antagonists can prevent cATP concentration increase caused by histamine while H1-R histamine antagonists are not as effective [42,43].

H1-R histamine antagonists stimulate tyrosinase activity and melanin accumulation, H2-R histamine antagonists decrease them. Enzyme activity and pigment synthesis are accompanied by cell proliferation inhibition of melanoma cells. CM effectiveness exceeds that of Ranitidinum. H2- R histamine antagonists do not stimulate melanogenesis in A375P and C32 melanoma cells, but they inhibit cell proliferation in both cell lines. Thus, H2-R histamine antagonists cause significant phonotypical changes in malignant melanoma cells in vitro [44,45].

Taking into account, accumulated data about the ability of H2-R histamine antagonists to effect melonogenesis process as well as the process of melanoma growth and cell metastasing, CM is used in a combined melanoma treatment [46]. CM has improved the effectiveness of cytokine therapy [47].

Application of combined therapy of CM, surgical treatment and biotherapy with autovaccine in the experiment on horses did not give positive therapeutic results [48]. However, taking into account the progress in the field of tumor biotherapy, the CM application as an adjuvant agent for melanoma treatment is considered to be expedient [49]. Promising data was aquired for malignant vulva melanoma treatment by means of combined therapy (surgical total excision of tumor), biotherapy with allogenic vaccine and CM immunotherapy [50]. It opens a possibility to CM application for a successful therapy of melanoma as immune regulating agent affecting the system and local immunological and physiological processes in the course of skin melanoma development.

The goal of this study is to investigate of the possibility of melanoma biotherapy effectiveness increase through modification of immune response in vivo using H2-receptor histamine antagonist (CM) as an adjuvant therapy agent for vaccinotherapy.

MATERIALS AND METHODS.

The research was performed in C57Bl6 male mice, 22-25 gram in weight, which were obtained from Animal Center (Russian Cancer Research Center, Moscow). Animals were treated according to the ethical guidelines of the Animal Centre, Russian Cancer Research Centre, Moscow. These mice were subcutaneously transplanted with melanoma cells. B16F10 mice melanoma cells were used which are similar to the immunophenotype of a human melanoma model. Growing tumors were measured twice a week, the tumor volume was calculated according to a formula: $V=D_{max} \times D_1 \times (D_{max}/2)$, where $D_{max}$ is maximum tumor diameter, $D_1$ is diameter perpendicular to the maximum one. The cells express a number of specific antigens and are characterized by the absence or low expression of MHC molecules. For the vaccinotherapy B16F10 BG melanoma GM-CSF- producing clone cells were used. They were exposed to γ-rays by Agat-R using 60Co at the experimental therapy clinic N.N. Blokhin Russian Cancer Research Center. Dose was 100 gray. Vaccine was preserved in liquid nitrogen. These manipulations were necessary for experimental reproduction of human clinical antitumor vaccination. We used Cimetidinum 200 mg (Pharmachim, Bulgaria). Antitumor effect was estimated by tumour growth inhibition and mice life span. Antimetastatic effect was evaluated by weight and number of the metastatic colonies in lungs according to standard techniques. Surgical operations were performed using gexanal anesthesia (100 mg/kg one time intra abdominal). Statistical analysis: the *P* values less than 0.05 were accepted as statistically significant.

RESULTS AND DISCUSSION.

48 hours later or with a delay after the formation of tumors (on 12-th days after transplantation B16F10 cells) mice were given CM *per os* five time in the doses of 25, 50 or 100 mg/kg. After completing the 5 days therapy CM antitumor effect was estimated by comparing average volumes of tumor nodes in control and treated mice, and as well as the dynamics during the experiment. After mice deaths that resulted from tumor growth progression the influence of CM (as a monotherapy) on melanoma mice life span and metastasis into lungs was estimated. The received results are shown a table 1.

**Table 1. Influence in various doses of CM to melanoma B16F10 growth in C57Bl6 mice.**

| Influence | Average life-time(days) | Life-time increasing (%) | Average tumor mass, gr | Tumor growth retarding (%) | Frequency of metastasis |
|---|---|---|---|---|---|
| Control group | 25,4±3,1 |  | 6,7±1,4 |  | 5/8 |
| CM (25 mg/kg)x5 (2-6 days) | 27,1±2,4 | 7 | 5,6±2,2 | 38 | 4/7 |
| CM (50 mg/kg)x5 (2-6 days) | 26,4±3,3 | 4 | 5,8±0,9 | 9 | 3/7 |
| CM (100 mg/kg)x5 (2-6 days) | 24,7±2,9 | -3 | 5,6±2,6 | 17 | 4/8 |
| CM (25 mg/kg)x5 (12-16 days) | 28,9±4,3 | 14 | 6,3±1,4 | 21 | 5/7 |
| CM (50 mg/kg)x5 (12-16 days) | 26,0±3,5 | 2 | 7,1±1,7 | 7 | 3/7 |
| CM (100 mg/kg)x5 (12-16 days) | 24,8±3,1 | -3 | 7,7±2,2 | 31 | 5/7 |

As we can see from the table 1, CM treatment during 5 days in given doses did not result in significant increase of average life span of treated mice in comparison with the control group at early and later stages of treatment. In the course of studies of CM effectiveness on growth and development of melanoma cells, intensity of metastasis, it was shown that at earlier timepoints (on the 2-6-th days of experiment) CM application in the dose of 25 mg/kg resulted in a slight inhibition of initial tumor node growth (tumor growth inhibition (TGI) = 38%). Dynamic observation of the average tumor node volumes in mice receiving CM in the dose 25 mg/kg has revealed that, the average tumor node volumes were less approximately by 1/3 in comparison with those in the control group. At the same time, tendency to increasing life span of the mice receiving CM in the dose of 25 mg/kg on the 12-16-th day of the experiment was noted. The average life span of these mice is 14% higher in comparison to the control (28.9±4.3 and 25.4±3.1 days correspondingly). Interestingly, a slight decrease of metastasization levels in the group of mice receiving treatment was noticed in comparison with the control. The results stipulate good assimilation of chosen doses and treatment methods.

In spite of the fact that these results cannot be called significant they are worth paying attention to and allow to consider CM as a perspective adjuvant therapy agent for melanoma treatment.

Based on the previous data, for the for the next experiments with combined treatment with antitumor vaccine the following dose was chosen: 25 mg/kg of CM (once a day during 5 days before vaccination; once a day during 5 days after vaccination and both (before and after) for different groups). Vaccination was performed 7 days before B16F10 tumor cells transplantation (preventive vaccination). The results are shown in Fig.1.

**Fig.1. CM effect on preventive vaccination.**

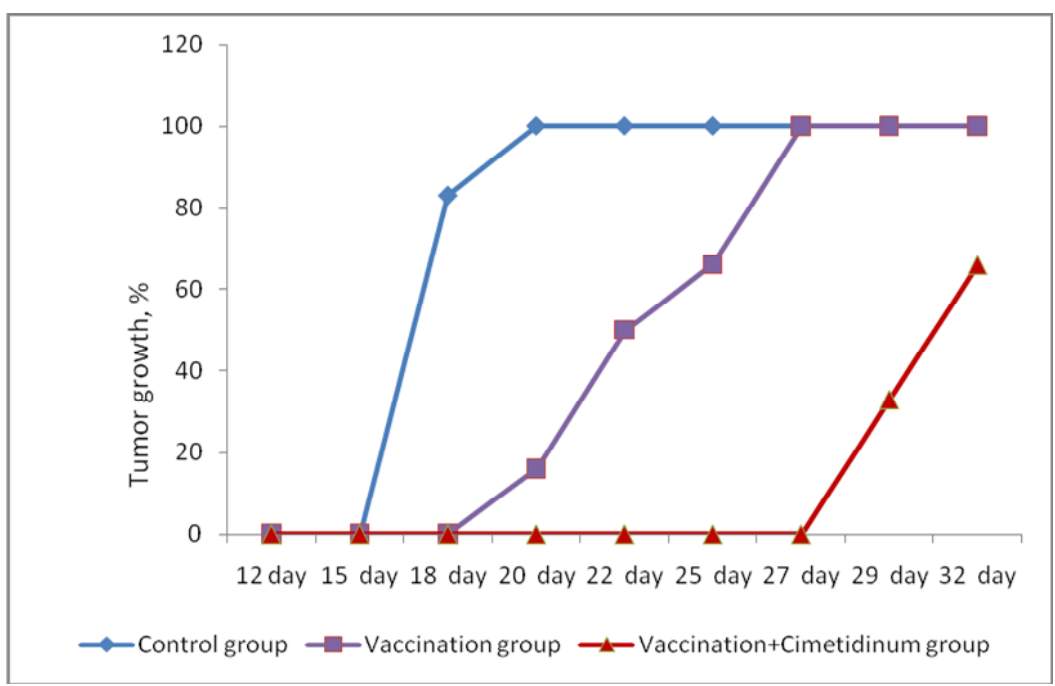

As we can see from the data in Fig.1, preventative vaccine application had a significant effect for reducing the growth of transplanted B16F10 cells. On day 29 of the experiment at the transplantation locus of B16F10 cells tumors developed only in 50% of mice. The volumes of these tumors were less 2-3 fold than in the control group.

Preventative vaccination along with CM treatment had more significant effect inhibiting tumor development. In such cases, palpable tumor nodes appeared in 66% of mice only on day 32of the experiment. In 33% of mice tumors did not develop during 60 days; at that time they were under observation. The average life span of mice receiving vaccine (without taking "cured" ones into account) along with CM treatment was 46.0±5.7 days (p<0.05) which is 35% more than those in the control group; if we take into account the mice that never developed tumors and have been observed for 60 days it is 68% more than in the control group. The results are shown in Fig.2.

**Fig.2. CM influence on mice life span using preventative vaccination.**

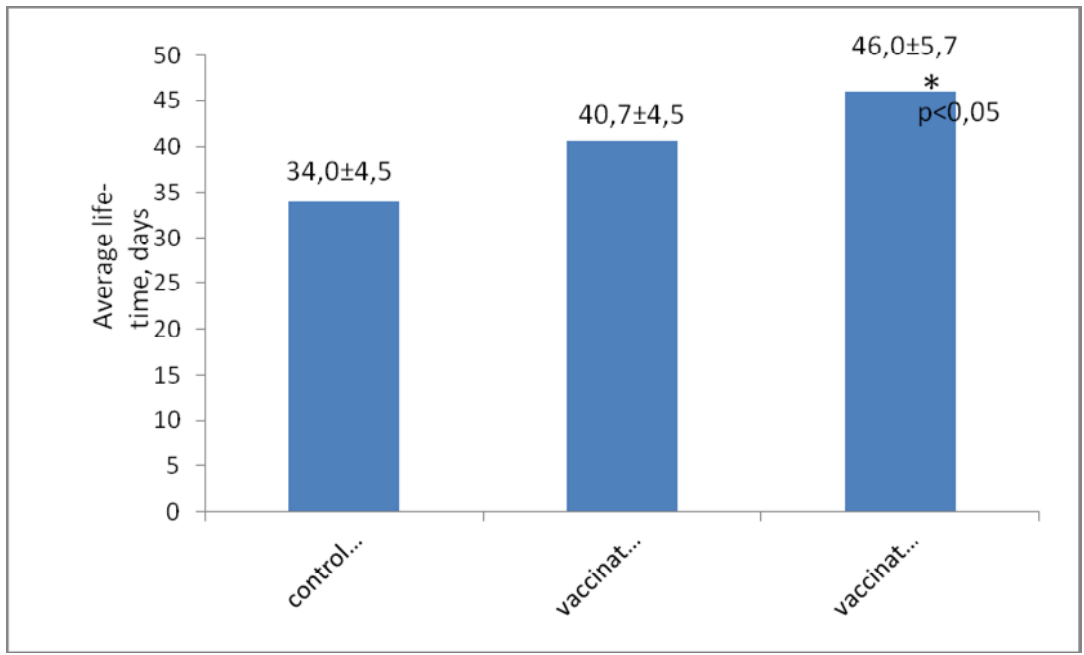

The results of using vaccination together with CM application for melanoma treatment (for treatment of already formed tumors – on the 7-th day after B16F10 cells transplantation) are shown in Fig.3.

**Fig.3. CM effect on vaccination for therapeutic treatment.**

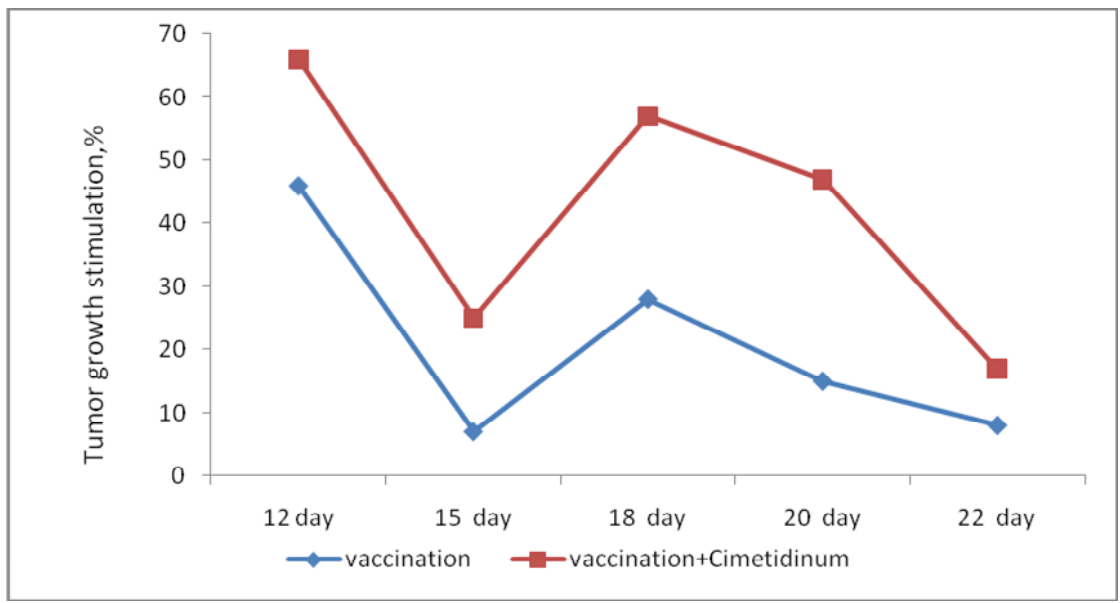

As it is seen from the data shown above in Fig.3 the application of vaccine together with CM treatment had significant stimulating effect on tumor node growth. At the same time we cannot exclude that in this case we deal with "false" stimulation of tumor growth stipulated by its high infiltration by immune cells and stromal cells that hyperproliferate in the tumor growth zone. Our idea is also supported by the data that show the same mice life span for the control group (24.0±2.3 days) and the one receiving vaccine (25.2±2.1 days), as well as the mice receiving vaccine along with CM treatment (26.0±2.4 days).

Because of the highly prominent immune suppression associated with high dose chemotherapy bone marrow transplantation is included into this experimental treatment method for the correction of this sideffect. Moreover, immune suppression associated with high-dose chemical therapy can decrease the level of premier lymphocytes, which lack cytotoxic activity. In this case application of immune therapy along with this therapy method could be more effective and have less sideffects.

In the present research it is shown that vaccine therapy, bone marrow transplantation and CM treatment did not increase the effectiveness of cytotstatics alone against the tumor nodes (life span increasing = 87%, p<0.05 in comparison with the control group). At the same time the effect of Cyclophosphamide alone in high dose was not effective enough in relation to metastases process in lungs. Only 43% of mice that have received Cyclophosphamide had developed metastases. CM therapy in combination with cytotstatics, bone marrow transplantation and vaccinotherapy as well as CM in high dose alone as chemical therapy did not result in increased therapeutic effectiveness. Results are shown in Table 2.

**Table 2. Combinatory therapy effectiveness (Cyclophosphamide, bone marrow transplantation, antitumor whole cell GM-CSF-producing vaccine, CM treatment).**

| Influence | Mice quantity died before control group (%) | Average life-time (days) | Life-time increasing (%) | Frequency of metastasis (%) |
|---|---|---|---|---|
| Control group | - | 22,7 ± 3,5 | - | 0 |
| Vaccination | 0 | 27,9 ± 3,7 | 23 | 14 |
| CM treatment | 14 | 23,4 ± 4,1 | 3 | 0 |
| Vaccination + CM treatment | 0 | 24,2 ± 2,0 | 6 | 14 |
| Cyclophosphamide + Marrow transplantation | 0 | 34,6 ± 4,2 | 52 (*p<0,05) | 0 |
| Cyclophosphamide + Marrow transplantation+ Vaccination | 0 | 40,9 ± 3,2 | 80 (*p<0,05) | 14 |
| Cyclophosphamide + Marrow transplantation+ Vaccination + CM treatment | 0 | 35,0 ± 4,6 | 54 (*p<0,05) | 16 |
| Cyclophosphamide + CM treatment | 0 | 32,6 ± 1,5 | 43 | 0 |
| Marrow transplantation | 28 | 23,4 ± 4,4 | 3 | 0 |
| Cyclophosphamide | 0 | 42,4 ± 6,1 | 87 (*p<0,05) | 43 |

Results from similar experiments using Cisplatin as chemical treatment are presented in Table 3. Application of Cisplatin with the following bone marrow transplantation and vaccination did not increase the effectiveness of Cisplatin alone similar to Cyclophosphamide application. At the same time application of Cisplatin in high doses increased the frequency of metastasis up to 57% in the treated mice group.

**Table 3. Combined therapy effectiveness (Cisplatine, bone marrow transplantation, antitumor whole cell GM-CSF-producing vaccine, CM treatment).**

| Influence | Mice quantity died before control group (%) | Average life-time (days) | Life-time increasing (%) | Frequency of metastasis (%) |
|---|---|---|---|---|
| Control group | - | 22,7 ± 3,5 | - | 16 |
| Vaccination | 0 | 27,9 ± 3,7 | 23 | 14 |
| Cisplatin + Marrow transplantation+ CM treatment | 0 | 28,9 ± 2,2 | 27 | 0 |
| Cisplatin + Marrow transplantation+ Vaccination + CM treatment | 14 | 25,6 ± 5,9 | 13 | 14 |
| Cisplatin | 14 | 29,1 ± 7,2 | 28 | 57 |
| Marrow transplantation | 28 | 23,4 ± 4,4 | 3 | 0 |

Acquired experimental results give us an opportunity to presume that it is necessary to correctly verify immune status for complex combined bio- and chemical therapies. It will allow creating better methods of combined therapy and increase antitumor and metastases activity.

Summary

1. Effectiveness of preventative antitumor whole cell GM-CSF-producing vaccine along with 5-time CM injection in the dose of 25 mg/kg is increased significantly.
2. Antitumor therapies such as high dose chemical therapy, bone marrow transplantation and CM (histamine H2 receptors inhibitor) treatment don't have a negative effect on treatment results.
3. Treatment with CM has a regulating effect on the type and kinetics of forming immune reaction, which allows recommending it for subsequent research as an adjuvant drug for antitumor whole cell vaccination.

*The research was made possible with financial support of Moscow Government in the framework of scientific and technical program "Development and implementation in medical practice new methods and ways of diagnostics and treatment of oncological and other diseases".*